\documentclass[twocolumn]{aa}
\usepackage{float}
\usepackage{graphicx}

\usepackage{lineno}
\usepackage{multirow}
\usepackage{subfigure}

\usepackage{scalerel}
\usepackage[]{xcolor}
\definecolor{citecolor}{HTML}{195D95}
\definecolor{green}{HTML}{20C29D}
\definecolor{BrickRed}{HTML}{C41010}

\usepackage[colorlinks=true,linkcolor=green,citecolor=citecolor,urlcolor=green]{hyperref}
\usepackage{amsmath,amssymb,latexsym}

\usepackage{newtxtext,newtxmath}
\usepackage{graphicx}
\usepackage{grffile}
\usepackage{lineno}
\usepackage{subfigure}
\usepackage{color}
\usepackage{amsmath,amssymb}
\usepackage{listings}
\usepackage[T1]{fontenc}
\usepackage{letltxmacro}

\definecolor{seagreen}{HTML}{20C29D}
\definecolor{BrickRed}{HTML}{C41010}

\def\vFv{\nu F_{\nu}}

\newcommand\given[1][]{\:#1\vert\:}

\begin{document}
\title{Is Spectral Width a Reliable  Measure of GRB Emission Physics?}

\author{J. Michael Burgess \inst{1,2} }

\institute{Max-Planck-Institut fur extraterrestrische Physik, Giessenbachstrasse 1, D-85748 Garching, Germany \\
  \email{jburgess@mpe.mpg.de}\label{mpe}
  \and Excellence Cluster Universe, Technische Universit{\"a}t M{\"u}nchen, Boltzmannstra{\ss}e 2, 85748 Garching, Germany\label{ec}
}
\date{}

\abstract{ The spectral width and sharpness of unfolded, observed GRB
  spectra have been presented as a new tool to infer physical
  properties about GRB emission via spectral fitting of empirical
  models. Following the tradition of the 'line-of-death', the spectral
  width has been used to rule out synchrotron emission in a majority
  of GRBs. This claim is investigated via reexamination of previously
  reported width measures. Then, a sample of peak-flux GRB spectra are
  fit with an idealized, physical synchrotron model. It is found that
  many spectra can be adequately fit by this model even when the width
  measures would reject it. Thus, the results advocate for fitting a
  physical model to be the sole tool for testing that model. Finally,
  a smoothly-broken power law is fit to these spectra allowing for the
  spectral curvature to vary during the fitting process in order to
  understand why the previous width measures poorly predict the
  spectra. It is found that the failing of previous width measures is
  due to a combination of inferring physical parameters from unfolded
  spectra as well as the presence of multiple widths in the data
  beyond what the Band function can model.}


\maketitle

\section{Introduction}
Catalogs of GRB observations contain spectra fit to the canonical Band
function \citep{Band:1993} which consists of two power laws that are
exponentially connected
\citep{Greiner:1995,Briggs:1999,Goldstein:2012,Gruber:2014,Yu:2016aa}. They
are additionally fit with other empirical photon models when the Band
function does not provide an acceptable fit. Historically, empirical
approaches to characterizing GRB spectra have focused on the Band
fitted low-energy power law slope, $\alpha$, from which conclusions
are drawn about the physical process producing the observed emission
\citep{Crider:1997,Preece:1998}. These studies find that a fraction,
$\sim 1/3$, of GRB spectra cannot be explained by the simplest
so-called slow-cooled synchrotron emission models and disfavor the
more preferred (on account of radiative efficiency) fast-cooled
synchrotron models \citep{Sari:1998,Beniamini:2013}. This has often
been referred to as the 'line-of-death' problem.

Further investigation of the spectra, aided by fitting physical
synchrotron models to the data \citep{Burgess:2014}, confirmed that
many GRB spectra cannot be fit by fast-cooled electron synchrotron
spectrum because the spectral width of the data was too narrow for
this model \citep[see however][]{Uhm:2014aa, Zhang:2016}. Yet, it was
found that some GRBs whose spectra violated the line-of-death were
able to be fit with slow-cooled synchrotron models directly. This
hinted that using the empirical Band function to characterize the
physical origin of GRB spectra can be misleading.

Recently, new empirical tools have been introduced in an attempt to
characterize the observed, prompt gamma-ray spectra of GRBs
\citep{Axelsson:2015,Yu:2015}. Motivated by the claims of
\citet{Beloborodov:2013aa} that synchrotron emission is too broad for
the observed data, these works take the unfolded empirical spectra
from the GRB catalogs and characterized them by an auxiliary quantity,
the spectral width, in an attempt to measure the broadness of the
observed spectra. Both \citet{Axelsson:2015} and \citet{Yu:2015}, then
compare these observed widths with the widths of physical spectra and
arrive at the conclusion that a large fraction of GRB spectra are
inconsistent with synchrotron emission.

Such empirical procedures are powerful tools in astronomy. Without
much effort or the need for computationally expensive physical models,
the community can quickly categorize thousands of observations and
provide tests for theoretical predictions from which models can be
rejected. For this reason, many studies have begun adopting the width
as a tool to advocate for photospheric emission \citep{Ahlgren:2015,
  Iyyani:2015aa, Iyyani:2016aa, Vurm:2016, Bharali:2017}. Therefore,
these empirical tools must fully incorporate the properties of the
observed data. The typical approach of post-processing unfolded,
fitted GRB spectra introduces a bias; the inferred properties of the
post-processing are influenced by properties of the functional form of
the already-fitted model, and lose information that was contained in
the raw count data. This is to say, that measuring the width of fitted
Band functions does not directly measure the width inherent in the
data. Herein, a different approach is taken to measuring the width of
GRB spectra in order to incorporate properties of the folded data into
empirical inferences. By modeling the width directly in the data
during the fitting process, any bias introduced by the Band function's
natural width is reduced.

Even with the use of more predictive physical measures, the process is
simply a substitute for the growing field of physical model fitting
\citep[e.g.][]{Burgess:2014,Ahlgren:2015,Zhang:2016}. Thus it is now
possible to evaluate the physical predictions of empirical measures
directly. If an observed GRB can be fit with a physical model that
would have been rejected by an empirical measure, then this empirical
measure must be disregarded. The current paradigm of GRB spectral data
modeling allows us to fit physical models directly to data, reject
those models when necessary, and develop better theoretical
predictions.


This article is organized into three main sections. First a review of
the approaches to measuring the width developed in
\citet{Axelsson:2015} and \citet{Yu:2015} (Section
\ref{sec:review_width}). Next, a sample of GRB spectra are fit with a
physical synchrotron spectrum and an evaluation of the quality of the
fit compared to the predictions of the empirical approaches is made
(Section \ref{sec:synchrotron}).  Finally, a method for measuring the
width of the spectra directly in the data by fitting a sample of GRB
peak-flux spectra is employed (Section \ref{sec:fit_width}).

\section{A Review of GRB spectral widths}
\label{sec:review_width}
Two different approaches to measuring the width or sharpness of GRB
data were undertaken by \citet{Axelsson:2015} and
\citet{Yu:2015}. Following \citep{Beloborodov:2013aa},
\citet{Axelsson:2015} define the width as the logarithmic ratio of the
energies at the full width half maximum (FWHM) spectra:

\begin{equation}
  W = \log_{10}\left(\frac{E_2}{E_1} \right)
\end{equation}
\noindent
and \citet{Yu:2015} defined a sharpness angle ($\theta$) at the $\vFv$
peak between two normalized fluxes at their respective normalized
energies. With these definitions, both works define limits of
different emission mechanisms in their respective measurement spaces
as shown in Table \ref{tab:physical_measures}. These mechanisms
include a Planck function, single-particle synchrotron (SPS),
synchrotron from a Maxwellian distribution of electrons (MS), and
synchrotron from a power law distribution of electrons (PLS) with
electron indices of either $p=2$ or $p=4$. In each approach, it was
found that a majority of the data cannot be explained by synchrotron
emission.

\begin{figure}
  \subfigure[]{\includegraphics[scale=1]{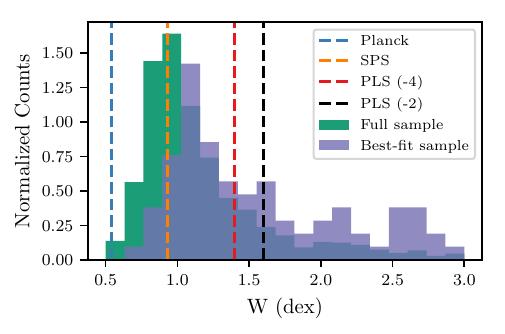}}
  \subfigure[]{\includegraphics[scale=1]{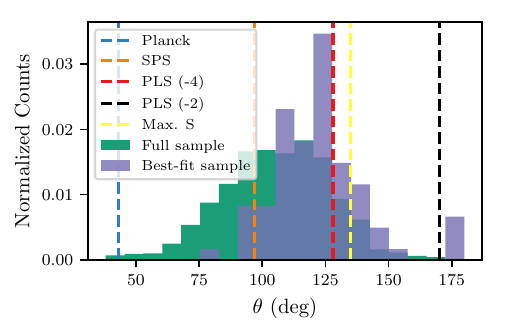}}
  \caption{Distributions of $W$ (top) and $\theta$ (bottom) for the
    entire sample and the best-fit sample.}
  \label{fig:hists}
\end{figure}

\begin{table}
  \begin{tabular}{cccccc}
    Width Measure
    & Planck &SPS & MS  & PLS
                          (-4)  & PLS (-2) \\
    \hline
    \hline
    $\theta$ (degrees) & 43 & 97 & 135 & 128 & 170 \\
    $W$ (dex) & 0.54 & 0.93  & 1.4  & 1.4 & 1.6 \\
  \end{tabular}
  \caption{Derived Physical Widths}
  \label{tab:physical_measures}
\end{table}

The sample selection in \citet{Axelsson:2015} included GRBs from the
Gamma-ray Burst Monitor (GBM) onboard the Fermi Gamma-ray Space
Telescope \citep{Meegan:2009}. The authors used all Band fits from the
GBM catalog peak-flux spectral catalog regardless of which photon
model best-fit the spectrum or if the Band function resulted
  in a failed maximum-likelihood fit. A cut was applied to the data
requiring the low- and and high-energy power laws of the Band function
($\alpha$ and $\beta$ respectively) be $\alpha >-1.9$ and
$\beta <-2.1$. Herein, this analysis is replicated and then a further
cut requiring that the best fitting spectrum as determined in the
catalog be either the Band function or a smoothly-broken power law
(SBPL) is applied. This eliminates spectra that may include Band
parameters from failed fits due to a simpler function such as the
exponentially-cutoff power law (CPL) or power law (PL) having been
recorded as the best-fit. The width and the sharpness angle are
computed from each observation in both the full and best-fit
samples. The software used to compute the width and sharpness angle is
released for the purpose of replication.
\footnote{\href{https://github.com/grburgess/width\_calculator}{https://github.com/grburgess/width\_calculator}}

Figure \ref{fig:hists} shows that that when the cuts for best-fit
spectrum are applied, the distributions shift to broader spectra or
away from thermal spectra and towards optically-thin synchrotron
spectra. \citet{Goldstein:2012} note that with increasing
signal-to-noise in the peak-flux spectra, the photon models shift from
simple models such as the PL and CPL to more complex models such as
the Band function or SBPL. This tentatively indicates that spectra are
best fit by these simpler functions due to lack of photon statistics
and not due to intrinsic physical reasons. The simpler functions are
intrinsically narrower than the Band function and SBPL. Therefore,
including spectra that were best fit by simpler (narrower) functions
and not the Band function in the sample artificially leads to a bias
towards narrower spectra. It is noted that \citet{Yu:2015} computed
the spectral sharpness on time-resolved spectra using the best-fitting
model of each observed spectrum of the GBM time-resolved catalog
\citep{Yu:2016aa}.

\begin{figure*}
  
  \subfigure[]{\includegraphics{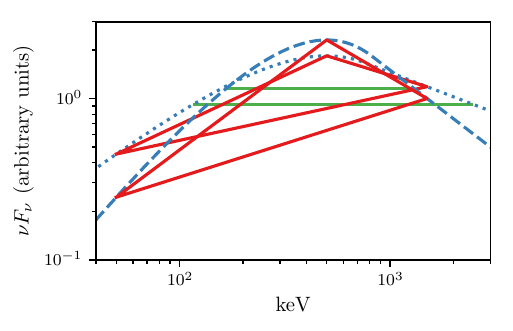}}\subfigure[]{\includegraphics{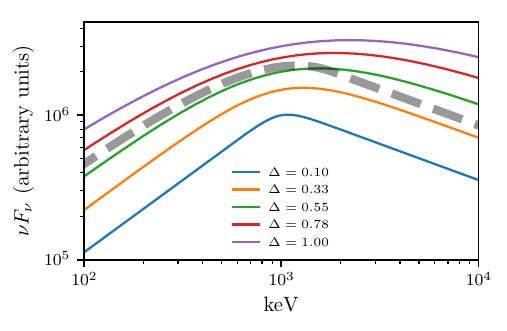}}
  \caption{(left) An illustration of how $W$ (green lines) and
    $\theta$ (via the red triangles) are realized on two different toy
    Band functions. (right) An example of how the SBPL varies with
    $\Delta$. For comparison, a Band function is superimposed in black
    dashed lines.}
  \label{fig:toy}
\end{figure*}

\begin{figure}
  
  \includegraphics{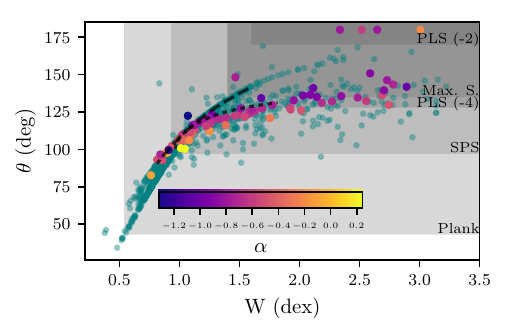}
  \caption{The $W-\theta$ plane from the full GBM catalog (teal) and
    the subsample of GRBs best fit by the Band function or
    a SBPL (purple-yellow). The boxes are the allowed regions for the
    corresponding physical emission mechanisms. The color of the
    subsample corresponds to the low-energy index ($\alpha$) of the
    spectral fit. The two black dashed lines demonstrate how the
    $W-\theta$ plain evolves for fixed $\beta$ (big dashes) and fixed
    $\alpha$ (small dashes).}
  \label{fig:w-theta}
\end{figure}

The two width measures differ in their prediction for what types of
spectra are viable. Figure \ref{fig:toy} gives a toy example of how
the measures look in $\vFv$-space. Examining the $W-\theta$ plane,
Figure \ref{fig:w-theta} shows the full sample against the GBM
best-fit sample as well as the regions allowed for different
models. Over-plotted are the relation between $W$ and $\theta$ with
$\beta=-2.25$ and $\alpha \in \{-1.5,0\}$ as well as $\alpha=-0.8$ and
$\beta \in \{-2.25,-4 \}$. Interestingly, the best-fit sample follows
a different trend than the full sample corresponding to the
fixed-$\alpha$ curve. There is also a tentative correspondence of
$\alpha$ and width with softer $\alpha$ values corresponding to larger
width. It is clear that when a cut is not applied to the
  catalog fits to ensure that no failed fits are included, the width
  measures move away from the a clustering at "non-synchrotron"
  allowed values. Nevertheless, as pointed out in each work, the
  standard synchrotron models are strongly rejected by the width
  measures.

\section{Synchrotron emission}
\label{sec:synchrotron}
Instead of employing empirical measures to infer if synchrotron can
fit the data, let us now fit the data with a synchrotron model. A
power law synchrotron model has been implemented into 3ML via {\tt
  astromodels}\footnote{\href{https://github.com/giacomov/astromodels}{https://github.com/giacomov/astromodels}}
following the method of \citet{Burgess:2014} and described in Appendix
\ref{sec:sm} except that the Maxwellian part of the electron
distribution is not included. In the spirit of open software, the
model is made publicly available for use with
3ML\footnote{\href{https://github.com/grburgess/synchrotron\_models}{https://github.com/grburgess/powerlaw\_synchrotron}}. The
model has three free parameters: a normalization, the electron
spectral index, and an energy scaling parameter (the magnetic field
strength) proportional to the peak of the $\vFv$ spectrum. While it is
numerically expensive to compute, it is functionally less complex than
the four parameter Band function and five parameter SBPL thus making
it less flexible.

\subsection{Sample selection}
\label{sec:sample_selection}
The GBM has observed over 2000 GRBs with cataloged spectral parameters
readily available
online\footnote{\href{https://heasarc.gsfc.nasa.gov/W3Browse/fermi/fermigbrst.html}{https://heasarc.gsfc.nasa.gov/W3Browse/fermi/fermigbrst.html}}. However,
we must re-fit these data for this work. All spectral data used
consist of 128 channel, time-tagged event (TTE) data obtained from the
Fermi Science Support Center (FSSC). GRBs with cataloged parameters
that were detected before 2017 and with best-fit peak-flux spectra of
either Band or SBPL are used. These criteria result in a sample of 91
GRBs. Some catalog entries in the FSSC database had invalid response
matrices and were discarded from the sample as it is important to use
the exact responses that were used to compute the cataloged
spectra\footnote{These errors involved RSP2 files that did not have
  valid time coverage intervals appropriate for the published
  peak-flux intervals.}. Next, a cut on significance over background
of 30$\sigma$ was introduced to have a bright sample. A requirement
that at least two Sodium Iodide (NaI) detectors in addition to one
Bismuth Germinate (BGO) detector have acceptable viewing geometry of
the GRB as denoted in the catalog further reduces the sample size. No
selection on previously fitted spectral parameters was made to
eliminate biasing the sample. Additionally, GRBs with Fermi Large Area
Telescope (LAT) data are cut as they may include additional spectral
features such as high-energy cutoffs. The final selections resulted in
a sample of 44 GRBs.

Using the information provided in the GBM catalog, detectors,
background and peak-flux intervals are selected to appropriately match
with the selections used to produce the catalog. It was required that
some background selections be modified as the ones specified in the
online catalog occasionally contained on-source intervals. With these
selections, the backgrounds were fitted with a series of polynomials
of varying order via an unbinned Poisson likelihood and the best one
was chosen via a likelihood ratio test (LRT). The modeled background
count estimation and Gaussian error were extrapolated into the source
interval as described in \citet{Greiner:2016kv}. For source intervals,
the 1.024 s peak-flux intervals denoted in the catalog were selected
to minimize the effects of spectral evolution as well as to keep the
properties of the sample close to those which were used in Section
\ref{sec:review_width}.

\subsection{Spectral fitting procedure}
\label{sec:fit_procedure}
For spectral analysis, the Multi-Mission Maximum Likelihood
framework\footnote{\href{https://github.com/giacomov/3ML}{https://github.com/giacomov/3ML}}
\citep[3ML][]{Vianello:2015aa} is used. The likelihood for the data is
a Poisson-Gaussian likelihood to account for the Poisson nature of the
total counts and the Gaussian nature of the modeled background
\citep{Arnaud:1996vl}.  3ML allows for both maximum likelihood (MLE)
and Bayesian posterior simulation (BPS) via a variety of optimization
or sampling algorithms. For this study, BPS was chosen via the {\tt
  emcee} \citep{Foreman-Mackey:2013} algorithm to fit the data in the
sample. The fitting procedure involves two steps. First, MLE is used
to find a starting point for the BPS. For MLE, the MINUIT
\citep{James:1975} optimization algorithm is used.  With the MLE
starting point, the posterior is sampled using flat, uninformative
priors on all parameters\footnote{Log uniform priors are used on scale
  parameters and uniform priors for spectral indices for all fitting
  in this work.} To account for systematics in the GBM response
matrices, the total effective area of all detectors is scaled to the
brightest NaI detector by multiplicative constants. Instead of
uninformative priors, informative Cauchy priors centered at unity,
i.e., no correction, and width set to reflect the assumed 10\%
systematics in the GBM responses \citep{Bissaldi:2009} are used. The
use of a Cauchy prior rather than a Gaussian is due to its wider shape
around the mean reflecting the lack of knowledge about the systematics
within 10\%, but the belief that they are not too extreme.

Model comparison between the empirical functions used in Section
\ref{sec:fit_width} and synchrotron is not attempted because an
empirical function can always be designed to fit the data with more
predictability than a physical model. Moreover, it is invalid to treat
the Band function as a null hypothesis against synchrotron emission as
it is not a hypothesis, a special case of synchrotron (a so-called
nested model), or part of a closed set of models which are known to
include the true data-generating process \footnote{This is known as
  the $\mathcal{M}$-closed model comparison scenario
  \citep{Vehtari:2012hq} under which LRTs and Bayes factors are a
  valid statistical test. This is explicitly not the case herein.}.In
fact, this is the goal of empirical models. Instead, model checking of
the synchrotron fits is performed via posterior predictive checks
(PPC) which allows us to see if the observed data look plausible under
the posterior predictive distribution. The details of the procedure
are discussed in Appendix \ref{sec:ppc}.

\subsection{Synchrotron Results}
\label{sec:synch_results}

\begin{figure}
  \subfigure[]{\includegraphics{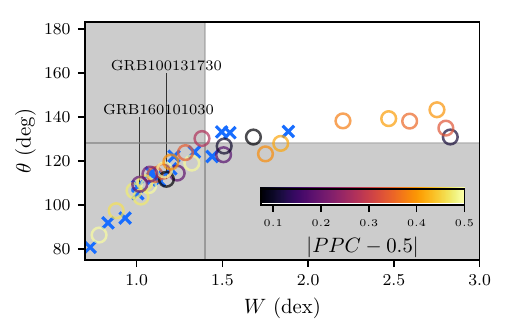}}
  \subfigure[]{\includegraphics{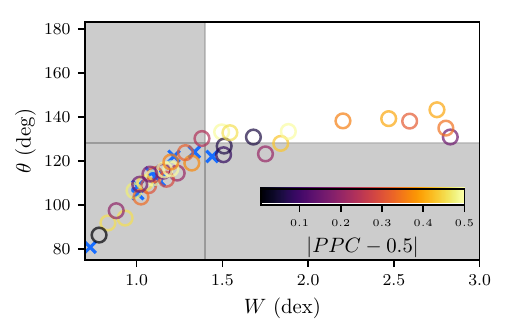}}
  \caption{The $W-\theta$ plane with synchrotron PPCs (top) and Band
    PPCs (bottom). Blue X's indicate extreme poor fits and color
    indicate deviation from 0.5, i.e., darker colors indicate better
    fits. The grey shaded regions are regions that would be excluded
    by synchrotron for an electron distribution of $p=-4$.}
  \label{fig:real_wt}
\end{figure}

\begin{figure*}
  \subfigure[]{\includegraphics{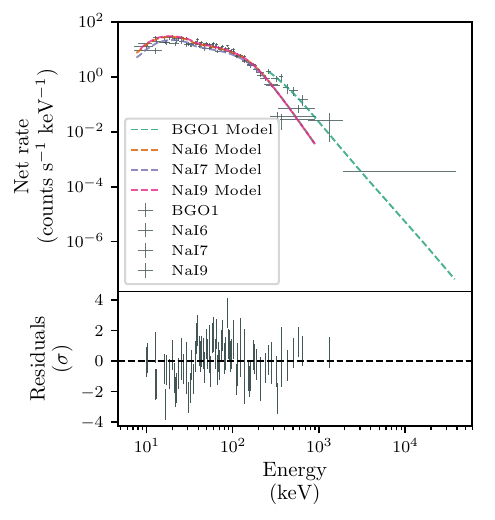}}\subfigure[]{\includegraphics{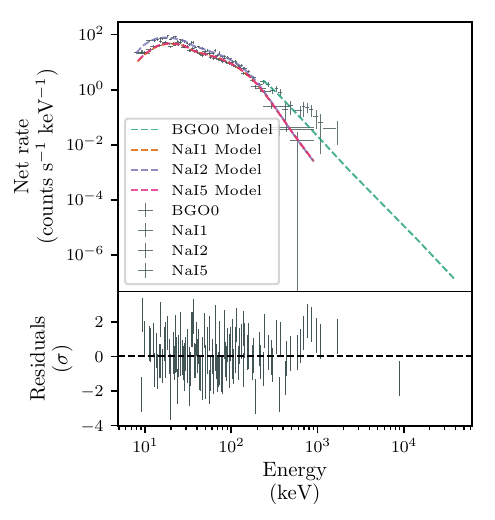}}
  \caption{The folded count spectra of two synchrotron fits (GRB
    100131730 (left) and GRB 160101030 (right)) that had acceptable
    PPCs but would have been rejected via spectral width and sharpness
    angle as indicated in Figure \ref{fig:real_wt}. The solid lines
    are the folded model through each GBM detector in the fit while
    the grey points indicate the raw count data.}
  \label{fig:synch_counts}
\end{figure*}

In order to directly compare a success fit of the synchrotron model to
the inference of the width measure, all data were also fitted by the
Band function allowing for the Band derived width to be computed. This
allows the fits to synchrotron to be displayed in the $W-\theta$ plane
with their PPC values in Figure \ref{fig:real_wt} (a). Both width
measures are computed from the Band function fit of the data. There
are some spectra that lie in the excluded regions that have extremely
poor PPC values as indicated by the blue X's; however, several fits
lie in the excluded region that can be well described by
synchrotron. We demonstrate two of these fits in Figure
\ref{fig:synch_counts} which had similar PPC values as those fitted
with the Band function. This explicitly demonstrates that models with
very different shapes in photon space can be statistically similar in
detector count space thus making empirically derived model assessment
procedures very uninformative. Similar results were recently shown in
\citet{Vianello:2018il}. This is likely due to the Band function not
properly modeling the inherent shape of the data and hence resulting
in an misleading $W$ or $\theta$ value. Therefore, empirical width
measures fail to accurately predict if synchrotron is a viable
spectral model for the data and hence cannot be used with the purpose
for which they were designed. Synchrotron fit parameters are displayed
in Appendix \ref{sec:synch}. Figure \ref{fig:bad_ppc} shows fits to
both synchrotron and the Band function where the PPC for synchrotron
was very bad and for the Band function acceptable. The pattern in the
residuals for the synchrotron fit is not dissimilar to that of Figure
\ref{fig:synch_counts} (a). Naively, this would indicate that both
fits should have bad PPCs. This is precisely why visual inspection of
residuals which measures only a single point in the posterior can be
misleading. PPCs have the advantage of measuring the deviation from
the data across the entire posterior.

Goodness of fit via any method should be regarded with caution because
one never has access to the true model. Moreover, it is preferable to
compare physically motivated models to each other and chose the one
which provides the best predictability of the data similar to what we
have done with the empirical models. Nevertheless, PPCs for the Band
function fits are computed and displayed in Figure \ref{fig:real_wt}
(b). We can see that many of the Band function fits also do not
accurately model the data according to the chosen PPC criterion,
though the number of poor Band fits was smaller than poor synchrotron
fits. The poor Band fits can be due to any number of issues such as
unmodeled detector systematics like the K-edge non-linearity at
$\sim 32$ keV \citep{Bissaldi:2009,Goldstein:2012}. Therefore, I
conclude that while synchrotron nor the Band function provide
universally adequate fits, synchrotron does fit some spectra which
would be ruled out by the width measures.

\begin{figure*}
  \centering
  \subfigure[]{\includegraphics{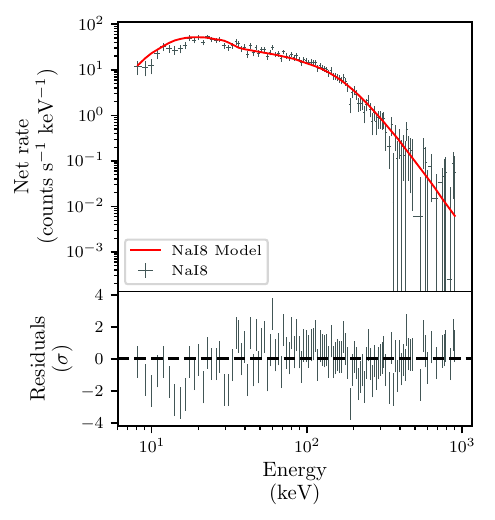}}\subfigure[]{\includegraphics{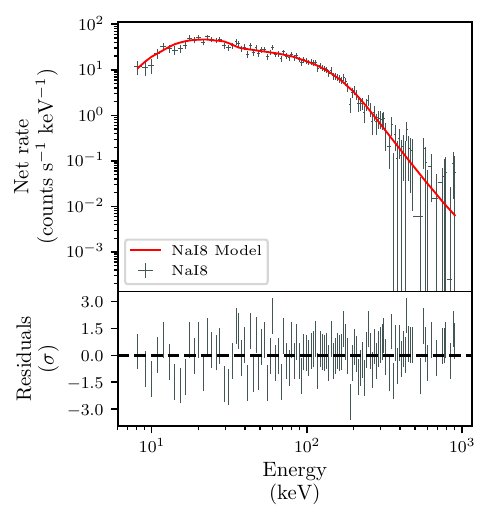}}
  \caption{The fit of synchrotron (left) and Band (right) to one of
    the detectors from GRB 160521385. Here, the PPC tail probability
    indicates that the synchrotron fit is very poor while the Band
    function accurately fits the data. A pattern can be seen in the
    residuals of the synchrotron fit. }
  \label{fig:bad_ppc}
\end{figure*}

\section{Fitting for the width}
\label{sec:fit_width}

It is important to understand why the previous empirical width
measures too conservatively reject synchrotron emission as shown in
Section \ref{sec:synch_results}. Thus, the data are fitted with an
SBPL of the form:
\begin{equation}
  \label{eq:sbpl}
  F_{\gamma}(\varepsilon) = A\left(\frac{\varepsilon}{\varepsilon_{\rm piv}}\right)^{b} 10^{(a-a_{\rm piv})}
\end{equation}
\noindent
where
\begin{align}
  \label{eq:sbpl2}  
  a=m\Delta \log\left(\frac{e^{q} +e^{-q}}{2}\right),\;& a_{\rm piv} = m\Delta \log\left(\frac{e^{q_{\rm piv}} +e^{-q_{\rm piv}}}{2}\right)\\
  q=\frac{\log_{10}(\varepsilon/\varepsilon_{\rm break})}{\Delta},&\;q_{\rm piv}=\frac{\log_{10}(\varepsilon_{\rm piv}/\varepsilon_{\rm break})}{\Delta} \\
  m=\frac{\beta - \alpha}{2},&\;b=\frac{\beta + \alpha}{2}\text{.}
\end{align}
\noindent
Here, $\varepsilon_{\rm break}$ is the break energy in keV,
$\varepsilon_{\rm piv}$ is the pivot energy, $\alpha$ and $\beta$ are
the low- and high-energy spectral indices respectively, and $\Delta$
is the break scale in decades of energy \citep[for a similar use of a
SBPL, see][]{Ryde:1999}. The proxy for the width of the spectral data
will be $\Delta$ as it is a measure that is optimized during the
fitting process and thus contains information about the width of the
folded data. Figure \ref{fig:toy} demonstrates how the function
changes as a function of $\Delta$ for a range covering the
distribution found from fitting the data. The intent of this fitting
is \emph{not} to introduce a new measure of width in the literature,
but rather to see if the Band function is adequately modeling all
features present in the data.

\subsection{Fit Results and Model Selection}

\begin{figure*}
  \subfigure[]{\includegraphics{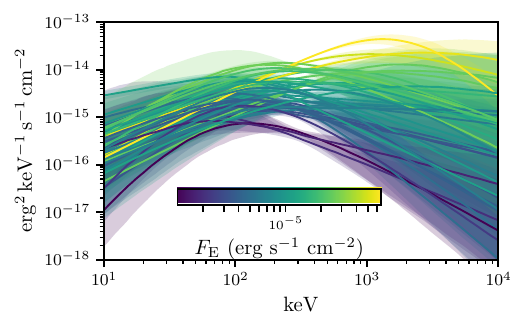}}\subfigure[]{\includegraphics{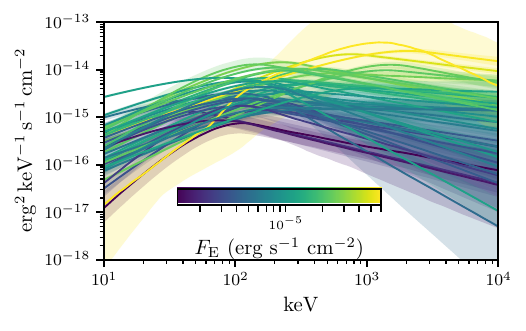}}
  \caption{The $\vFv$ spectra and 1$\sigma$ contours of the SBPL
    (left) and Band (right) fits. The color corresponds to the 10
    keV-4 MeV integrated energy flux ($F_{\rm E}$). The SBPL fits
      result in broader or smoother curvature around the $\vFv$ peak.}
  \label{fig:vfv_spectra}
\end{figure*}

Every GRB spectrum is fit to both the SBPL and Band function so that a
comparison between the models can be made. The complexity of the SBPL
function can result in local minima regardless of the optimization
scheme; therefore, optimizing on a logarithmic grid of $\Delta$ and
spectral normalizations results in a more robust result. Figure
\ref{fig:vfv_spectra} illustrates the results of the fits to the two
different functions. It can be seen that the SBPL allows for
  more posterior variance below the $\vFv$ peak due to its freedom to
  vary its width during the fit. Model selection via the likelihood
ratio test (LRT) between the Band function an SBPL is not possible due
to the fact that they are not nested functions. Additionally, for the
empirical functions used, the aim is to assess whether a richer model
is required to describe the data. The deviance information criteria
(DIC) can be used to judge which model provides the best
predictability of the data (See Appendix \ref{sec:dic}). Table
\ref{tab:results} details the results of the fits. Of the spectra fit,
all but one are best described by the SBPL, i.e., positive
$\delta_{\rm DIC}$ in Table \ref{tab:results}.

The ability to fit a width parameter in the data indicates that the
spectra have a variety of inherent widths rather than a single natural
width of the Band function. Any change in the empirically
  measured width of the Band function is related to a change in
spectral indices only. This variety is not captured by the Band
function and hence, widths derived from the Band function can be
systematically biased. It is noted that in GBM spectral catalogs, an
SBPL is also fit to the spectra but its $\Delta$ is always fixed to
0.3 for historical reasons related to the bandpass of BATSE
\citep{Kaneko:2006}.

\begin{figure}
  \includegraphics{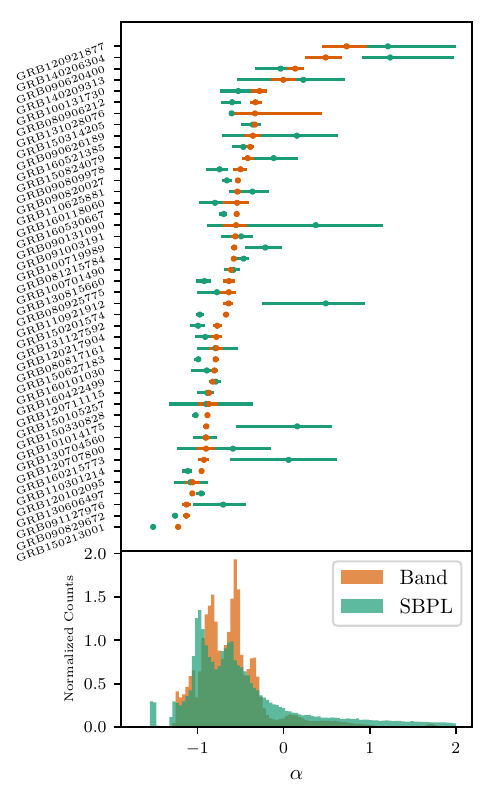}
  \caption{The distributions of the low-energy index ($\alpha$) from
    the Band and SBPL fits. The error bars represent the 0.68 highest
    posterior density intervals. There are systematic differences
    between the values of $\alpha$ due to the free curvature of the
    SBPL. The lower histograms are produced from the combined marginal
    distributions of the fits to fully incorporate the uncertainty in
    the fits.}
  \label{fig:spectral_difference}
\end{figure}

Another interesting feature of the SBPL fits is the different values
of the measured $\alpha$ values. The distribution of $\alpha$ from the
SBPL is shifted to softer values with a tail extending to hard values
(see Figure \ref{fig:spectral_difference}). Noticeably larger
uncertainty on SBPL $\alpha$'s is due to the additional freedom in the
curvature. Physical inferences coming from empirical models are
dependent on the spectral shape from which they are derived. The long
standing paradigm that the Band function's $\alpha$ should be used to
infer physical spectra as well as the newly proposed limits of width
measures do not hold if the Band function is not the best-fit to the
data.

\subsection{The smoothly-broken power law and synchrotron}
\label{sec:data_width}

\begin{figure}
  
  \includegraphics{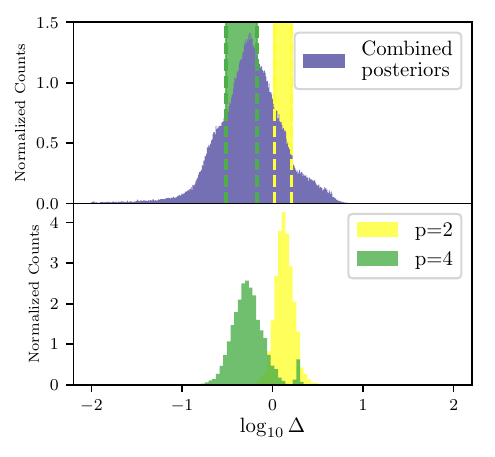}
  \caption{The distribution of $\Delta$ from the fitted GRB spectra
    (top). Uncertainty on $\Delta$ is included via the marginal
    posteriors of each fit. The lower panel displays the marginal
    distributions of $\Delta$ for simulated synchrotron spectra with
    $p=2,4$. The 68\% credible regions for these fits are superimposed
    on the full distribution. This demonstrates that the theoretical
    widths from pure synchrotron emission are consistent with the
    widths derived from the sample.}
  \label{fig:delta}
\end{figure}

Let us now examine the proxy for the width in the data, $\Delta$, and
its relation to synchrotron emission. To incorporate the uncertainty
on $\Delta$ into the full sample distribution, the full marginal
distributions from all fits are combined into a single distribution in
Figure \ref{fig:delta}. The distribution is unimodal with a tail
extending to narrow widths. Hints of substructure are visually
apparent in the distribution, but are likely an artifact of small
sample size.

A simple power law synchrotron model from
\citet{Baring:2004,Burgess:2014} was used to create synthetic count
spectra using the GBM response matrices from a GRB in the
sample. These synthetic spectra were then fit via BPS to a SBPL to
estimate the values of $\Delta$ for different electron power law
distributions. However, several factors influence the value of
$\Delta$ beyond the shape of the electron distribution alone; most
notably, the number of counts at high-energy in the synthetic
spectra. This makes it difficult to set a hard limit on which values
of $\Delta$ correspond to various synchrotron scenarios\footnote{Both
  \citet{Axelsson:2015} and \citet{Yu:2015} calculate their respective
  limits in photon space rather than count space. Both works use Monte
  Carlo methods to calculate reportedly small errors on their
  respective measures.}. Nevertheless, examining the distribution of
$\Delta$ expected from synchrotron with electron power law indices
$p=2,4$ is necessary to follow the previous investigations into the
spectral width. These limits are displayed in Figure \ref{fig:delta}
both as the full marginal distribution from the BPS and as their
respective 0.68 credible regions. The peak of the observed $\Delta$
distribution coincides with the SBPL-fitted $\Delta$'s of the
synthetic synchrotron spectra when $p=4$. Thus, the distribution of
widths from real data are marginally consistent with the limits
derived from pure synchrotron emission. This is notably in contrast to
the conclusions derived in previous studies providing further evidence
that the natural width of the (less predictive) Band function is
biasing the previously derived widths.

\begin{figure}
  \subfigure[]{\includegraphics{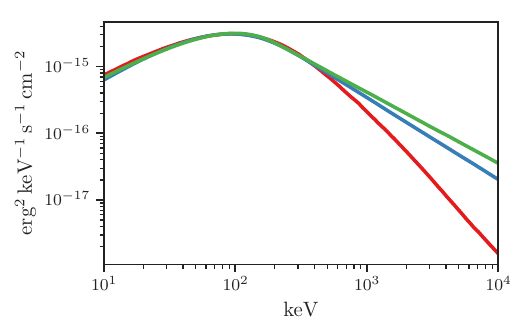}}
  \subfigure[]{\includegraphics{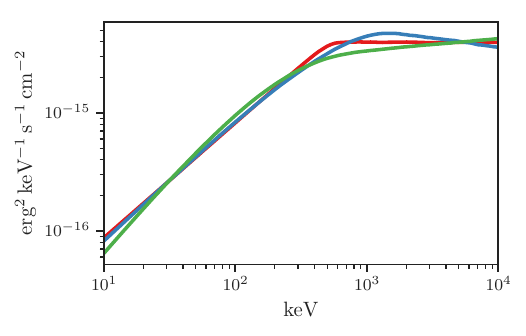}}
  \caption{Two examples of how the SBPL (blue) and Band (red) function
    fit a synthetic synchrotron spectrum (green). On top, an example
    where the SBPL models the synchrotron (W=-1.4, $\theta$=128)
    emission adequately while the Band function is too narrow and
    artificially softens $\beta$ to compensate. On bottom, both the
    SBPL and Band functions are poor approximations of the true
    synchrotron spectrum (W=-1.6, $\theta$=170) and hence would result
    in poor empirical inference about the true underlying mechanism.}
  \label{fig:compare_phys}
\end{figure}

While these results are promising for synchrotron emission, it is
worth examining their weaknesses. The difficulties of reconciling
empirical models with physical spectra presents us with problems even
when the more flexible SBPL is used to fit and characterize the
spectra. Consider Figure \ref{fig:compare_phys} which shows the fitted
Band and SBPL functions to simulated synchrotron spectra. In the first
case (Figure \ref{fig:compare_phys} (a)) the SBPL accurately models
the spectra, in the second case (Figure \ref{fig:compare_phys} (b))
the SBPL overestimates the peak energy of the synthetic model and
poorly models the non power law behavior of spectrum at low
energies. In each case, the Band function fails to accurately model
the synthetic spectrum. For these reasons, even though $\Delta$ serves
as a better proxy for the width due to its ability to model the width
directly in the data, it should not be used to set quantitative limits
or inferences for the true underlying emission mechanisms in the data.

\section{Discussion}
It has been demonstrated that the empirical width measures derived in
\citet{Axelsson:2015} and \citet{Yu:2015} fail to predict when GRB
data cannot be fit with a synchrotron spectra. Many of the spectra
could be adequately fit with this model in regardless of the value of
the spectral width derived from the Band function. These measures are
derived from an empirical models that can poorly represent synchrotron
emission. This was previously highlighted in \citet{Lloyd:2000aa}
where parameterized synchrotron models were fit directly to the data
and their asymptotic low-energy power law behavior was compared to
that of the Band function. Only a small subset of GBM peak-flux
spectra were examined and hence no physical conclusions about the
validity of the synchrotron model used herein can be drawn. Such
conclusions require examining the time-resolved spectra of individual
GRBs. Instead, it was assessed whether empirical width measures which
would have rejected synchrotron failed when synchrotron was actually
fit to the same spectra. Furthermore, the question of whether
synchrotron emission can be rejected when compared to other physical
models has not been posed in this work as there are no other publicly
available physical models to compare against. In reality, an emission
model like synchrotron possesses multiple widths which may change
non-linearly with the model parameters. This further exaggerates the
issues with using secondary inference methods in fitted model
space. This was not addressed directly in this study because the fact
that the width fails to accurately predict the underlying model in the
simple case is enough to demonstrate that fitting physical spectra is
the proper way forward in GRB emission studies.

In an attempt to understand the spectral width by including the width
of the data rather than the unfolded model, a sample of GRBs was
re-fit with a SBPL and the distribution of its break scale parameter
($\Delta$) was examined. With this measure, GRBs exhibit a variety of
inherent data widths and the majority of these widths are \emph{not}
inconsistent with synchrotron emission. While this approach is a more
appropriate empirical measure of the spectral width, it too suffers
from the problem that the SBPL does not always model the shape of
synchrotron emission properly. It is not entirely surprising that
these empirical measures do not serve as quantitative inferences for
physical models. \citet{Burgess:2014,Burgess:2015} showed that
synchrotron emission can fit GRBs that violate the 'line-of-death' and
that Band $\alpha$ values provide little insight into the presence of
blackbodies in GRB spectra. Thus, it is strongly suggested that the
fitting of physical models be performed to ascertain which model best
represents the data.

While model checking can provide a qualitative guide to the validity
of a spectra fit, other information should be used to fully justify
the use of a model. These information can include physically motivated
priors and predictions from time-evolving models such as those used in
\citet{Dermer:1999,Peer:2008,Bosnjak:2014}. For example,
\citet{Burgess:2016} used temporal predictions from
\citet{Dermer:1999} combined with synchrotron spectral fits to argue
for an external shock interpretation of GRB 141028A. Therefore,
without temporal or other information, it is important to not stress
the physical implications of this work. Empirical models provide no
such insight to physics other than assessing general features about
the data e.g., the total flux, peak $\vFv$ energy and the existence of
high-energy power laws. Deeper physical inferences from these
empirical models should be regarded with caution until verified by
complimentary analysis with physical emission models.

\section*{Supporting Material}
All processed GBM data files for use with XSPEC or 3ML as well as the
analysis files containing the parameter marginals (readable by 3ML)
are released for the purposes of replication. In addition, sample code
for constructing the models used is also released. All files can be
located here:
\href{https://dataverse.harvard.edu/dataset.xhtml?persistentId=doi:10.7910/DVN/BDC2GS}{doi:10.7910/DVN/BDC2GS}.

\section*{Acknowledgements}

I gratefully acknowledge fruitful discussion with Damien B\'egu\'e,
Giacomo Vianello, David Yu, Patricia Schady, and Thomas Kr{\"u}hler
concerning various aspects of this work.

\bibliographystyle{aa} \bibliography{bib}

\begin{table*}
\def\arraystretch{1.5}%
  \begin{tabular}{ccccccccc}
    GRB & $\alpha_{\rm Band}$ & $\alpha_{\rm SBPL}$ & $\Delta_{\rm SBPL}$ & $\delta_{\rm DIC}$ & $p_{\rm eff}^{\rm Band}$ & $p_{\rm eff}^{\rm SBPL}$ & $p_{\rm B}^{\rm Band}$ & $p_{\rm B}^{\rm synch}$ \\
    \hline
    \hline
    080817161 &  $-0.78_{-0.08}^{+0.08}$ &  $-0.79_{-0.22}^{+0.26}$ &  $0.53_{-0.53}^{+0.35}$ &      9.01 &      5.79 &     -1.51 &       0.184 &       0.144 \\
    080906212 &  $-0.33_{-0.07}^{+0.07}$ &  $-0.60_{-0.13}^{+0.10}$ &  $0.35_{-0.14}^{+0.14}$ &      6.10 &      6.52 &      5.43 &       0.196 &       0.004 \\
    080925775 &  $-0.63_{-0.10}^{+0.08}$ &  $-0.77_{-0.23}^{+0.15}$ &  $0.44_{-0.27}^{+0.21}$ &      5.27 &      5.62 &      2.92 &       0.172 &       0.150 \\
    081215784 &  $-0.58_{-0.02}^{+0.02}$ &  $-0.46_{-0.09}^{+0.06}$ &  $0.83_{-0.16}^{+0.12}$ &     19.71 &      6.91 &      5.14 &  $<10^{-3}$ &  $<10^{-3}$ \\
    090131090 &  $-0.55_{-0.16}^{+0.12}$ &   $0.37_{-1.26}^{+0.78}$ &  $0.76_{-0.44}^{+0.23}$ &      1.88 &     -6.96 &     -5.12 &       0.046 &       0.026 \\
    090620400 &   $0.14_{-0.10}^{+0.10}$ &  $-0.03_{-0.30}^{+0.20}$ &  $0.51_{-0.24}^{+0.15}$ &      9.40 &      5.99 &      1.55 &       0.042 &  $<10^{-3}$ \\
    090626189 &  $-0.35_{-0.10}^{+0.09}$ &   $0.15_{-0.86}^{+0.48}$ &  $0.90_{-0.52}^{+0.37}$ &     24.69 &      4.53 &    -20.26 &       0.712 &       0.310 \\
    090809978 &  $-0.50_{-0.08}^{+0.08}$ &  $-0.74_{-0.15}^{+0.10}$ &  $0.28_{-0.28}^{+0.12}$ &      6.62 &      5.98 &      3.38 &       0.658 &       0.384 \\
    090820027 &  $-0.53_{-0.03}^{+0.03}$ &  $-0.66_{-0.06}^{+0.06}$ &  $0.45_{-0.12}^{+0.07}$ &     25.51 &      5.91 &     -2.66 &       0.004 &  $<10^{-3}$ \\
    090829672 &  $-1.13_{-0.04}^{+0.04}$ &  $-1.26_{-0.04}^{+0.03}$ &  $0.21_{-0.12}^{+0.13}$ &      9.58 &      5.90 &      5.90 &       0.004 &  $<10^{-3}$ \\
    091003191 &  $-0.56_{-0.06}^{+0.06}$ &  $-0.49_{-0.24}^{+0.14}$ &  $0.79_{-0.41}^{+0.22}$ &     24.86 &      5.45 &    -14.41 &       0.412 &       0.318 \\
    091127976 &  $-1.13_{-0.05}^{+0.05}$ &  $-0.70_{-0.35}^{+0.27}$ &  $1.28_{-0.37}^{+0.28}$ &     24.09 &      5.95 &    -18.80 &       0.260 &       0.226 \\
    100131730 &  $-0.28_{-0.10}^{+0.09}$ &  $-0.53_{-0.21}^{+0.15}$ &  $0.35_{-0.23}^{+0.18}$ &      6.67 &      5.76 &      4.03 &       0.790 &       0.426 \\
    100701490 &  $-0.61_{-0.04}^{+0.04}$ &  $-0.58_{-0.11}^{+0.08}$ &  $0.71_{-0.27}^{+0.22}$ &      8.83 &      5.83 &     -2.53 &       0.338 &       0.156 \\
    100719989 &  $-0.57_{-0.04}^{+0.04}$ &  $-0.21_{-0.23}^{+0.19}$ &  $1.10_{-0.32}^{+0.27}$ &     33.40 &      5.90 &    -11.16 &       0.004 &  $<10^{-3}$ \\
    101014175 &  $-0.90_{-0.04}^{+0.04}$ &   $0.16_{-0.71}^{+0.40}$ &  $2.63_{-1.05}^{+0.76}$ &     83.79 &      5.27 &    -46.02 &       0.096 &       0.092 \\
    110301214 &  $-0.95_{-0.03}^{+0.03}$ &  $-1.11_{-0.07}^{+0.05}$ &  $0.43_{-0.09}^{+0.11}$ &     11.69 &      6.80 &      1.73 &  $<10^{-3}$ &  $<10^{-3}$ \\
    110625881 &  $-0.53_{-0.03}^{+0.04}$ &  $-0.36_{-0.27}^{+0.19}$ &  $0.80_{-0.30}^{+0.23}$ &     25.06 &      6.97 &     -9.94 &       0.128 &       0.006 \\
    110921912 &  $-0.64_{-0.06}^{+0.05}$ &   $0.49_{-0.74}^{+0.46}$ &  $2.10_{-0.63}^{+0.65}$ &     34.57 &      5.32 &    -14.51 &       0.066 &       0.084 \\
    120102095 &  $-1.06_{-0.09}^{+0.08}$ &  $-1.08_{-0.19}^{+0.21}$ &  $0.60_{-0.60}^{+0.45}$ &      9.97 &      3.38 &     -4.15 &       0.902 &       0.900 \\
    120217904 &  $-0.78_{-0.07}^{+0.06}$ &  $-0.91_{-0.11}^{+0.07}$ &  $0.37_{-0.20}^{+0.18}$ &      8.73 &      6.45 &      3.91 &       0.566 &       0.606 \\
    120707800 &  $-0.90_{-0.11}^{+0.10}$ &  $-0.59_{-0.66}^{+0.44}$ &  $0.84_{-0.70}^{+0.45}$ &     14.31 &      4.86 &     -9.81 &       0.176 &       0.158 \\
    120711115 &  $-0.86_{-0.04}^{+0.04}$ &  $-0.88_{-0.12}^{+0.08}$ &  $0.59_{-0.42}^{+0.37}$ &     18.55 &      5.76 &    -14.57 &       0.562 &       0.422 \\
    120921877 &   $0.73_{-0.29}^{+0.24}$ &   $1.21_{-0.31}^{+0.79}$ &  $0.69_{-0.22}^{+0.23}$ &     -5.04 &     -3.63 &      4.71 &       0.042 &  $<10^{-3}$ \\
    130606497 &  $-1.06_{-0.02}^{+0.01}$ &  $-0.95_{-0.06}^{+0.05}$ &  $1.20_{-0.23}^{+0.21}$ &     20.49 &      5.92 &     -1.13 &       0.040 &  $<10^{-3}$ \\
    130704560 &  $-0.90_{-0.04}^{+0.05}$ &  $-0.90_{-0.15}^{+0.13}$ &  $0.63_{-0.14}^{+0.11}$ &      3.14 &      4.60 &      5.47 &  $<10^{-3}$ &  $<10^{-3}$ \\
    130815660 &  $-0.63_{-0.07}^{+0.07}$ &  $-0.92_{-0.10}^{+0.07}$ &  $0.25_{-0.11}^{+0.09}$ &      7.52 &      6.41 &      5.34 &       0.002 &       0.002 \\
    131028076 &  $-0.33_{-0.24}^{+0.78}$ &  $-0.60_{-0.03}^{+0.03}$ &  $0.55_{-0.06}^{+0.06}$ &   1955.85 &   -169.51 &      6.02 &  $<10^{-3}$ &  $<10^{-3}$ \\
    131127592 &  $-0.77_{-0.04}^{+0.05}$ &  $-0.99_{-0.10}^{+0.08}$ &  $0.38_{-0.15}^{+0.12}$ &     16.32 &      6.60 &     -4.45 &       0.024 &       0.022 \\
    140206304 &   $0.49_{-0.24}^{+0.19}$ &   $1.24_{-0.32}^{+0.74}$ &  $0.73_{-0.16}^{+0.20}$ &      2.58 &     -3.03 &      2.96 &       0.506 &       0.002 \\
    140209313 &  $-0.00_{-0.14}^{+0.14}$ &   $0.23_{-0.77}^{+0.49}$ &  $0.77_{-0.51}^{+0.30}$ &     11.28 &      3.47 &     -4.71 &       0.162 &       0.026 \\
    150105257 &  $-0.87_{-0.13}^{+0.11}$ &  $-0.90_{-0.43}^{+0.54}$ &  $0.44_{-0.44}^{+0.34}$ &      5.47 &      2.67 &     -1.37 &       0.314 &       0.288 \\
    150201574 &  $-0.67_{-0.03}^{+0.04}$ &  $-0.97_{-0.05}^{+0.05}$ &  $0.26_{-0.06}^{+0.06}$ &     12.53 &      5.93 &      6.92 &  $<10^{-3}$ &  $<10^{-3}$ \\
    150213001 &  $-1.22_{-0.02}^{+0.02}$ &  $-1.51_{-0.02}^{+0.02}$ &  $0.20_{-0.04}^{+0.04}$ &     52.70 &      7.03 &      6.62 &  $<10^{-3}$ &  $<10^{-3}$ \\
    150314205 &  $-0.33_{-0.04}^{+0.04}$ &  $-0.36_{-0.13}^{+0.10}$ &  $0.56_{-0.15}^{+0.11}$ &     15.87 &      5.81 &      1.90 &  $<10^{-3}$ &  $<10^{-3}$ \\
    150330828 &  $-0.88_{-0.03}^{+0.03}$ &  $-1.02_{-0.04}^{+0.03}$ &  $0.27_{-0.11}^{+0.12}$ &     13.57 &      6.76 &      4.64 &       0.002 &  $<10^{-3}$ \\
    150627183 &  $-0.79_{-0.03}^{+0.04}$ &  $-0.99_{-0.05}^{+0.04}$ &  $0.27_{-0.10}^{+0.08}$ &     12.08 &      5.84 &      5.40 &       0.136 &       0.104 \\
    150824079 &  $-0.42_{-0.06}^{+0.07}$ &  $-0.11_{-0.35}^{+0.28}$ &  $1.10_{-0.40}^{+0.35}$ &     24.03 &      6.44 &     -8.87 &       0.290 &       0.108 \\
    160101030 &  $-0.80_{-0.04}^{+0.05}$ &  $-0.89_{-0.18}^{+0.11}$ &  $0.55_{-0.21}^{+0.14}$ &      6.46 &      6.73 &     -1.25 &       0.342 &       0.306 \\
    160118060 &  $-0.54_{-0.16}^{+0.13}$ &  $-0.79_{-0.19}^{+0.15}$ &  $0.22_{-0.22}^{+0.17}$ &      9.30 &      1.41 &     -4.00 &       0.844 &       0.834 \\
    160215773 &  $-0.92_{-0.07}^{+0.06}$ &   $0.06_{-0.69}^{+0.57}$ &  $2.96_{-1.29}^{+1.05}$ &     19.55 &      4.46 &     -7.91 &       0.862 &       0.880 \\
    160422499 &  $-0.82_{-0.02}^{+0.02}$ &  $-0.79_{-0.08}^{+0.06}$ &  $0.72_{-0.16}^{+0.14}$ &     50.64 &      6.80 &    -17.48 &  $<10^{-3}$ &  $<10^{-3}$ \\
    160521385 &  $-0.38_{-0.05}^{+0.04}$ &  $-0.47_{-0.13}^{+0.09}$ &  $0.58_{-0.12}^{+0.12}$ &     20.82 &      6.76 &     -4.74 &       0.700 &       0.034 \\
    160530667 &  $-0.54_{-0.03}^{+0.03}$ &  $-0.69_{-0.05}^{+0.04}$ &  $0.42_{-0.07}^{+0.06}$ &     17.84 &      6.97 &      6.47 &  $<10^{-3}$ &  $<10^{-3}$ \\        
  \end{tabular}
  \caption{Spectral Fitting Results}
  \label{tab:results}
\end{table*}

\appendix

\section{Synchrotron Modeling}
\label{sec:sm}
In order to fit synchrotron emission directly to the data, we will
assume a simple and pragmatic parameterization. More importantly, we
will assume the parameterization which captures the assumptions used
to derive the rejection criteria for the various width measures
adopted in previous works. The observed emission is assumed to come
from a power law distribution of electrons that have been accelerated
by an unspecified mechanism. Thus,

\begin{equation}
  \label{eq:4}
  n_{\mathrm{e}}(\gamma) \propto \gamma^{-p} \; \forall \gamma \ge \gamma_{\mathrm{inj}}
\end{equation}
\noindent

where $\gamma$ is dimensionless electron energy,
$\gamma_{\mathrm{inj}}$ the energy at which electrons are injected
with spectral index $p$. The electrons are assumed to not cool via
their radiation of synchrotron photons within a dynamical
time. Therefore, we simply compute the synchrotron emission of this
power law distribution by convolving it with the standard synchrotron
emission kernel \citep{Blumenthal:1970aa}. Therefore,

\begin{equation}
  \label{eq:7}
      n_{\gamma} \left(\varepsilon ; N, B, \gamma_{\mathrm{inj}}, \gamma_{\mathrm{max}},  p \right) = N  \int_{\gamma_{\mathrm{inj}}}^{\gamma_{\mathrm{max}}}  \mathrm{d} \gamma \; n_{\mathrm{e}} \left(\gamma; p\right) \Phi\left(\frac{\varepsilon}{\varepsilon_{\mathrm{crit}}(\gamma; B ) } \right)
\end{equation}
\noindent
where 
\begin{equation}
  \label{eq:5}
  \Phi\left( w\right) = \int_{w}^{\infty} \mathrm{d}x\; K_{5/3} \left(x \right)
\end{equation}
\noindent
and
\begin{equation}
  \label{eq:6}
  \varepsilon_{\mathrm{crit}} \left(\gamma ; B \right) = \frac{3}{2} \frac{B}{B_{\mathrm{crit}}} \gamma^2 \mathrm{.}
\end{equation}
\noindent

Here, $N$ is an arbitrary spectral normalization constant, $B$ is the
magnetic field strength, $B_{\mathrm{crit}} = 4 \cdot 10^{14}$ G, and
$\gamma_{\mathrm{max}}$ is the maximum electron energy which is set to
have the spectral cutoff of the photon model above the GBM energy
range.

With this simple parameterization, $B$ and $\gamma_{\mathrm{inj}}$ are
multiplicatively degenerate in setting the $\vFv$ peak of the
spectrum, thus, the choice is made to fix
$\gamma_{\mathrm{inj}} =10^5$ which is an arbitrary choice. This
implies that the value of $B$ found during fits is scaled and cannot
be interpreted physically other than setting the location of the
$\vFv$ peak. Therefore, there are three fitting parameters: $B$, $p$,
and the arbitrary spectral normalization $N$. The high-energy shape of
the photon spectrum is set by $p$ while the low-energy shape is that
of the synchrotron kernel. Note that this is the same functional
form of synchrotron that is used to derive the conditions for
rejecting synchrotron emission via the width or the line of death.

\section{Synchrotron Fit Parameters}
\label{sec:synch}

\begin{figure}
  \subfigure[]{\includegraphics[scale=1]{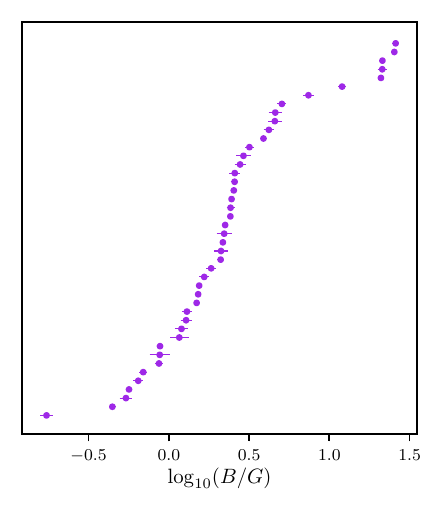}}
  \subfigure[]{\includegraphics[scale=1]{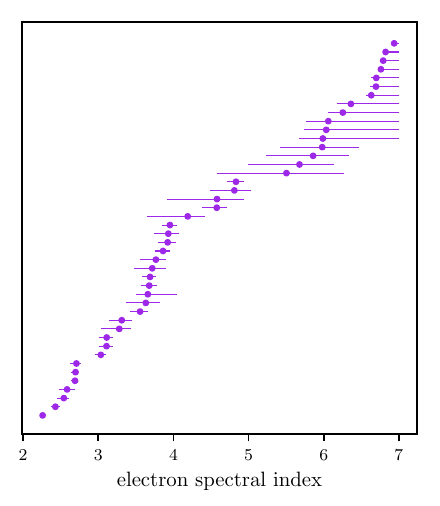}}
  \caption{The parameter distributions of $B$ and $p$ from the
    synchrotron fits along with their 68\% credible regions.}
  \label{fig:synch_param}
\end{figure}

Here I include parameter plots for the synchrotron fits for
reference. The synchrotron model in this work contains only two shape
parameters and a normalization. Figure \ref{fig:synch_param} display
the magnetic field strength and power law electron injection index
from the fits. The plots are ranked and display the 68\% highest
density posterior intervals obtained from the fit. The electron
spectral index is often found to be quite steep. This is not in
conflict with expectations from relativistic, oblique shock
acceleration theory \citep{Baring:2006aa}.

It is stressed that the value of the magnetic field strength value has
no scale unless assumptions about the emission region (emission
radius, time scale, etc.) are assumed as noted in \ref{sec:sm}. For
further discussion see \citet{Burgess:2014}.

\section{Deviance Information Criteria}

\label{sec:dic}
Model selection is one of the most difficult procedures in spectral
analysis. The use of reduced $\chi^2$ as a model rejection criterion
is not applicable to photon counting problems though it has previously
been employed in GBM spectral catalogs. The lack of Gaussian
likelihoods, generally non-linear models, unattained asymptotics of Wilk's
theorem \citep{Wilks:1938} and the generally non-nested models employed
\citep{Protassov:2002} violate a host of regularity
conditions required to apply simple hypothesis testing. Additionally,
the effective number of free parameters in a model is not necessarily
equal to the number of fitted functional parameters. Therefore, rather
than likelihood ratio tests, information criteria which seek to
quantify the predictive accuracy of a model can be more useful for the
current situation \citep[see however the technique of ][for an
approach to assessing non-nested likelihood ratio
tests]{Algeri:2016}\footnote{While Bayes factors and marginal
  likelihoods can also avoid the typical problems of the LRT, they are
  sensitive to the chosen prior distributions.}.

The Akaike information criteria (AIC) \citep{Akaike:1977} has recently
become common in X-ray spectral analysis as a model comparison tool
\citep{Zhang:2011,Buchner:2014}; however, it relies of point
estimates, an assumption of large number statistics, and only
penalizes model complexity by the number of free model parameters. The
deviance information criteria (DIC), uses the posterior mean, rather
than a point estimate and penalizes model complexity with the
effective number of free parameters which are a function of both the
data and the model \citep{Spiegelhalter:2002}.

\begin{equation}
  \label{eq:1}
  \text{DIC}= -2 \log \pi(y \given[] \hat{\theta} ) + 2 p_{\rm eff}
\end{equation}

\noindent
where $y$ are the data, $\hat{\theta}$ is the posterior mean and
$ p_{\rm eff}$ is the effective number of free parameters. The
effective number of free parameters is a function of both the model
and the data and can be negative if the posterior mean is far from the
mode \citep{Gelman:2014}. This allows for a model and data sensitive
measure of a model's data predictability.

\section{Posterior Predictive Checks}
\label{sec:ppc}
Assessment of a model's fit to data via posterior predictive checks
(PPCs) allows for incorporating information in the posterior into a
quantitative goodness of fit measure for future observations. The
usefulness of PPCs in X-ray spectral analysis has been demonstrated in
\citet{vanDyk:2004}. PPCs offer a guide to model assessment but are
simply a self-consistency check. In the current situation we lack other
physical models with which to check against. The posterior predictive
distribution is defined as

\begin{equation}
  \label{eq:2}
  \pi(y^{\rm rep} \given[] y) = \int d\theta \pi(y^{\rm rep} \given[] \theta) \pi(\theta \given[] y)
\end{equation}
\noindent

where $y$ are the data and $y^{\rm rep}$ are data replicated from a
the posterior and $\theta$ are the parameters. One way to assess the
lack of fit of data to this distribution is a tail-area probability
known as the posterior p-value:
\begin{equation}
  \label{eq:3}
  p_{\rm B} = \pi(T(y^{\rm rep},\theta) \ge T(y,\theta)\given[] y)\text{.}
\end{equation}
\noindent

Here, $T$ is a test statistic. For this work, 500 replicated spectra
are produced from the simulated posterior and $T(y,\theta)$ is
defined as the likelihood value for the given parameter. We compare
these test statistics to that of the actual data to arrive at a
measure of goodness of fit. A fit that adequately models the data
should have $p_{\rm B}\sim 0.5$ \citep{Gelman:2013}. Thus, we define a
so-called good fit as being $\vert p_{\rm B} - 0.5 \vert \sim 0$.

Model assessment and comparison is an ongoing and active part of
statistical research. Further details can be found in
\citep{Vehtari:2012hq}.

\label{lastpage}
\end{document}